\documentclass[
 reprint,
 amsmath,amssymb,
 aps,
 superscriptaddress
]{revtex4-2}

\usepackage[colorlinks=true]{hyperref}
\usepackage{graphicx}
\usepackage{dcolumn}
\usepackage{bm}

\hypersetup{allcolors=blue}

\begin{document}

\title{Radiation Damage Cascades in Fullerite Using Molecular Dynamics}

\author{Ethan P. Turner}
\email{ethan.turner@postgrad.curtin.edu.au}
\affiliation{Department of Physics and Astronomy, Curtin University, Perth, WA 6845, Australia}

\author{Paolo Raiteri}
\affiliation{School of Molecular and Life Sciences, Curtin University, Perth, WA 6845, Australia}

\author{Nigel A. Marks}
\affiliation{Department of Physics and Astronomy, Curtin University, Perth, WA 6845, Australia}

\date[]{}

\begin{abstract}
Molecular dynamics is used to study radiation cascades in solid C$_{60}$ under ambient conditions.
Simulations are performed for Primary Knock-On Atom (PKA) energies from 0.1 to 1~keV, and cascades are sampled over many PKA directions to collect  statistics. 
Energies and forces are described using the Environment Dependent Interaction Potential for carbon paired with the Ziegler-Biersack-Littmark potential for short-range interactions, and cascade behaviour is characterized by tracking kinetic energy, hybridization and bond connectivity as a function of time.
Compared to most materials, fullerite exhibits an unusual radiation response due to weak thermal transfer between C$_{60}$ molecules leading to a thermalization phase lasting hundreds of picoseconds. 
The cascades damage the C$_{60}$ molecules and link them together, and a linear relation is found between the number of cross-linked molecules and the number of new sp$^3$ atoms. 
The threshold displacement energy computed is 18 eV, in agreement with experiments.
\end{abstract}

\maketitle

\section{Introduction}

The discovery of C$_{60}$ in 1985~\cite{KROTO-C60} was a landmark moment in carbon science, leading to a new allotrope of carbon and laying the groundwork for the later discoveries of nanotubes~\cite{IIJIMA-CNT} and graphene~\cite{NOVOSELOV-GRAPHENE}. 
Looking back, the discovery of fullerenes was arguably the starting point for the field of nanotechnology.
Solid C$_{60}$, known as fullerite, was first synthesized in 1990~\cite{KRATSCHMER-FULLERITE} and was extensively studied in the subsequent decade. 
Under ambient conditions, fullerite consists of rapidly rotating C$_{60}$ molecules whose centre-of-mass sits on a face-centred-cubic lattice. 
At lower temperatures and higher pressures the molecules develop orientational order and simple-cubic symmetry, while at liquid nitrogen temperatures a glass is formed; see Refs.~\cite{SUNDQVIST-1999,SUNDQVIST-2021} for reviews.

In recent years there has been significant interest in using fullerite as a precursor for novel phases such as ultra-hard carbons~\cite{SHANG-ULTRAHARD-a-C}, covalently-bonded monolayer networks~\cite{MEIRZADEH-FULLERENE-NETWORK,HOU-FULLERENE-NETWORK}, and ordered porous carbons~\cite{PAN-LRPOROUSCARBONS}. 
Attractive aspects of fullerite include numerous reactive sp$^2$ sites, ease of chemical processing, and the absence of elemental contaminants such as oxygen, nitrogen and hydrogen.
Many different methods have been developed to cross-link the C$_{60}$ cages, including pressure~\cite{SUNDQVIST-2004}, intercalation of metals~\cite{TANAKA-METAL-INTERCALATION}, photo-induced polymerization (primarily UV~\cite{RAO-PHOTOINDUCED-UV-OPT,ONOE-PHOTOINDUCED-UV,ALVAREZ-PHOTOINDUCED-UV} or optical light~\cite{HASSANIEN-PHOTOINDUCED-OPT,ZHOU-PHOTOINDUCED-OPT}) and electron irradiation~\cite{ZHAO-ELECTRON-POLYMERISED}. 
Ion bombardment is another route by which fullerite can be cross-linked, but the vast majority of studies have been experimental, typically using high-energy protons (0.1--2~MeV)~\cite{MATHEW-PROTON-BOMBARDMENT,MUSKET-PROTON-BOMBARDMENT} or swift heavy-ions (30~keV to 200~MeV)~\cite{PALMETSHOFER-ION-BOMBARDMENT-keV,ZAWISLAK-ION-BOMBARDMENT-keV,NARAYANAN-ION-BOMBARDMENT-MeV}.

Computer simulations of C$_{60}$ irradiation effects have been relatively sparse, with most focusing on the use of C$_{60}$ as a projectile~\cite{SMITH-1996-C60Proj-Si,WEBB-1999-C60Proj-Graphite,WEBB-2001-C60Proj-Graphite,POSTAWA-2003-C60Proj-Ag,CHANG-2004-C60Proj-Diamond,KRANTZMAN-2008-C60Proj-Dia-Gra,PARUCH-2009-C60Proj-organics,RUSSO-2009-C60Proj-Ag,RESTREPO-2011-C60Proj-MetalOrganic,CZERWINSKI-2013-C60Proj-Polystyrene,WEBB-2011-C60Proj-Fullerite,CZERWINSKI-ARN-C60-FULLERITE}. Only a handful of simulations~\cite{HOBDAY1997-AR-FULLERITE,SHINYA2009-C02-FULLERITE,WEBB-2011-C60Proj-Fullerite,CZERWINSKI-ARN-C60-FULLERITE} have explored C$_{60}$ as a target material, starting in 1997 with Hobday \emph{et al}.\ \cite{HOBDAY1997-AR-FULLERITE} who modelled 300--1000~eV argon projectiles impacting onto fullerite thin films. They measured the sputter yield and also described the collision cascade, which involves cross-linking of molecules via sp$^3$ connections and the partial transition to amorphous carbon at collision sites. Several other groups also modelled the bombardment of fullerite thin films: Shinya \emph{et al}.~\cite{SHINYA2009-C02-FULLERITE} used carbon-dioxide clusters (1 to 20 molecules), Webb \emph{et al}.~\cite{WEBB-2011-C60Proj-Fullerite} used C$_{60}$ itself, and Czerwinksi \emph{et al}.~\cite{CZERWINSKI-ARN-C60-FULLERITE} used both argon clusters (18 to 2500 atoms) and C$_{60}$. These studies examined cratering, sputter yield and the effect of different targets (diamond, graphite, benzene, octane). None of these works involved bulk fullerite, and the cascade dynamics were generally not the primary focus.

In this work, we present molecular dynamics (MD) simulations of cascades in bulk fullerite.
To the best of our knowledge, this is the first study of this type.
We use carbon primary knock-on atoms (PKA) as the projectile, initialized with kinetic energies of 0.1--1~keV. 
For each energy, statistics are collected using 25 initial directions distributed uniformly across the surface of a sphere.
The ballistic phase completes within 0.5~ps and is characterized by computing the formation and breaking of bonds, cascade extent, and the kinetic energy maximum. 
Kinchin-Pease~\cite{KP} analysis of the threshold displacement energy finds a value in good agreement with other studies. 
The thermalization phase is extremely long-lived due to the weak inter-molecular forces and is followed for 300~ps, far longer than is necessary in most solids.
Cluster analysis during thermalization is used to extract the temperature of clusters, the distribution of kinetic energy and changes in bonding/cross-linking between molecules.
For context, we compare our findings throughout to our previous studies of cascades in graphite~\cite{CHRISTIE-GRAPHITECASC} and diamond~\cite{BUCHAN-DIAMONDCASC}.


\section{Methodology}

Atomic interactions are described using the Environment Dependent Interaction Potential (EDIP) for carbon~\cite{EDIP}.
EDIP is a highly transferable potential and has been extensively applied to diverse carbon systems including graphite~\cite{CHRISTIE-GRAPHITECASC,Graphite-EDIP-2023}, diamond~\cite{BUCHAN-DIAMONDCASC,Dia-EDIP-2021}, nanotubes~\cite{CNT-EDIP-2010,CNT-EDIP-2012}, carbon onions~\cite{CO-EDIP-2007,CO-Dia-EDIP-2012}, glassy carbon~\cite{GC-EDIP-2018,GC-EDIP-2019}, nanoporous carbon~\cite{CDC-EDIP-2017,CDC-EDIP-2018} and amorphous carbon~\cite{a-C-EDIP-2003, ta-C-EDIP-2005}. 
For close interactions (less than $\sim$1~\AA), EDIP is combined with the Ziegler-Biersack-Littmark (ZBL) potential~\cite{ZBL} using Fermi-type scaling functions that smoothly merge both potentials without inflection points~\cite{CHRISTIE-GRAPHITECASC}. 
Calculations are performed using an in-house Fortran code which includes a variable timestep algorithm \cite{VT} that simplifies the cascade simulations by providing good energy conservation without requiring user input.

The calculations begin with an energy minimization of a single fullerene molecule. 
Minimized C$_{60}$ within EDIP contains a single bond length of 1.49~\AA, in contrast to the experimental observation of two different bond lengths: 1.40 and 1.45~\AA~\cite{YANNONI-1991-NMR}.
The inability of EDIP to predict long and short bonds arises from an atom-centered bond-order term which cannot capture the subtleties of bond conjugation in C$_{60}$.
Despite this limitation, the strong transferability of EDIP, in particular its realistic energy barriers for making and breaking bonds, makes it well-suited to radiation damage studies.
The longer bond length produced with EDIP results in a C$_{60}$ diameter of 7.378~\AA\ and a cubic lattice parameter of 15.0~\AA, which are both $\sim$5\% larger than the respective values observed in experiment~\cite{HEDBERG-DIAMETER-1991,FLEMING-1990-XRD}.  

Fullerite structures are constructed in a simple cubic box with side lengths ranging from 45 to 150~\AA\, and the center-of-mass of each molecule rests on a face-centered-cubic lattice site. 
The simulation cell contains between 6,480 and 240,000 atoms and periodic boundary conditions are applied.  
The structures are equilibrated for 5~ps at 300~K, driving rotation of the molecules to mimic the orientational disorder seen experimentally at room temperature. 

The PKA in each cascade is initialized at one of five energies: 0.1, 0.25, 0.5, 0.75 and 1.0~keV. 
A representative set of cascades is sampled by using 25 different directions for each PKA energy and each direction is derived by evenly spacing points across the surface of a unit sphere (a 25-point solution of the Thomson problem~\cite{THOMSON-PROBLEM-1904}). 
A total of 125 simulations are performed following the cascade dynamics for 300~ps, and numerical integration is implemented using the Verlet algorithm. 
Thermostatting is not applied during the simulations and the box sizes chosen are sufficiently large to prevent cascade self-interaction.

Identifying defects in the fullerite cascades is not straightforward.
In many materials one can simply compare initial and final coordinates, in conjunction with a suitably chosen vacancy radius.
This method does not work for fullerite as molecular rotation from the initial thermalization and subsequent passage of the cascade would result in the entire system being identified as a defect.
We instead employ a neighbour list algorithm to monitor formation and breaking of bonds. 
Displacements are defined as those atoms which break all of their initial sp$^2$ bonds, and the cascade length is the maximal distance between any two displacements. Bonds are defined with a cutoff of 1.85~\AA, and the same cutoff is used to compute coordination fractions (sp, sp$^2$, sp$^3$).
The neighbour list is also used for cluster analysis to quantify cross-linking of the C$_{60}$ molecules by the cascade. We validated our code by comparing with a similar tool in the OVITO software package~\cite{OVITO}. OVITO was also used to visualise the cascades and create the videos in the supplementary material. 

A variety of time-dependent datasets are extracted from the cascades, and all are statistically analysed in a similar manner. 
For the first 0.5~ps, corresponding to the initial ballistic phase, a fine bin width of 0.01~ps is used for the kinetic energy maximum and 0.05~ps is used for the bond statistics and displacements. 
Up to 5~ps, a larger bin width of 0.1~ps is used for the kinetic energy maximum and a value of 0.25~ps for bond statistics and displacements.  
Beyond 5~ps, corresponding to the extended thermalization, a coarse bin width of 10~ps is employed for all statistics. 
Data in each bin is analysed to compute the mean across the 25 PKA directions, and unless otherwise specified, all error bars indicate the standard error in the mean.


\section{Results}

\begin{figure*}[t]
    \centering
    \includegraphics[width=\linewidth]{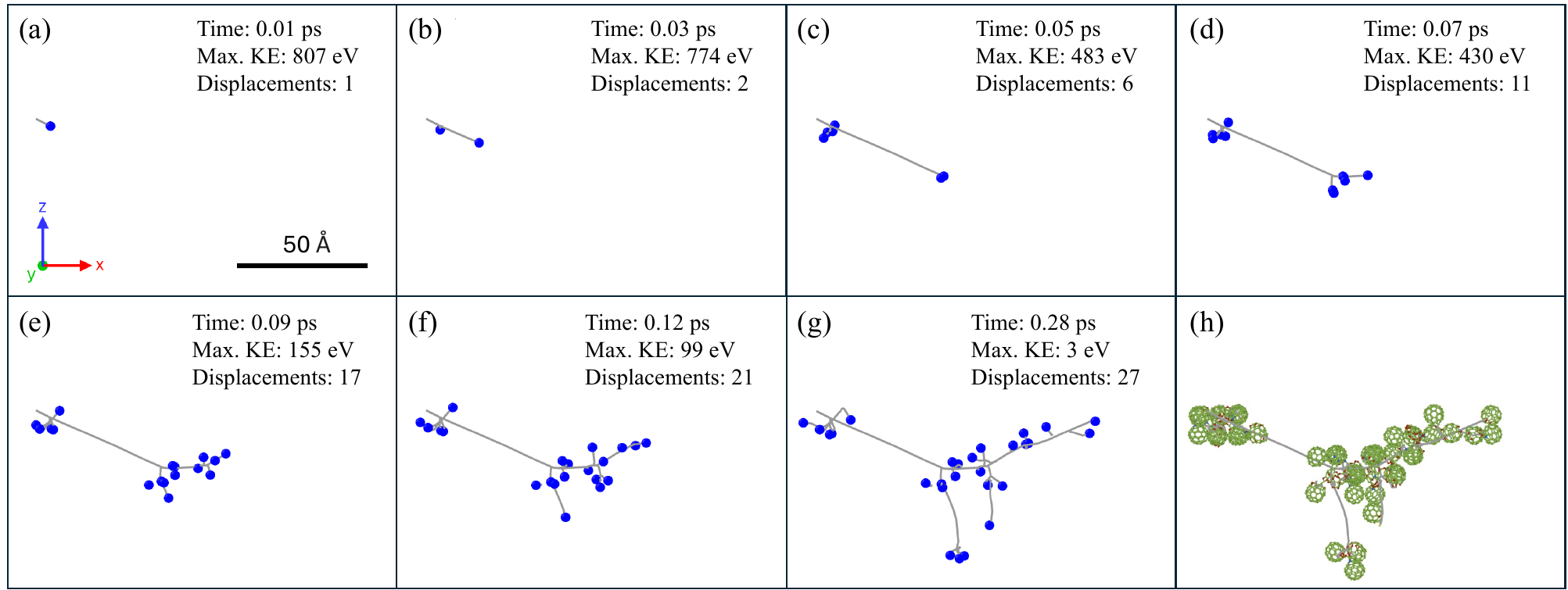}
    \caption{(a-g) Time series of the ballistic phase in a typical 1~keV cascade in fullerite. Blue spheres indicate displacements and grey lines show their trajectories. (h) Same time as (g), but using cluster analysis. Clusters with exactly 60 atoms are omitted, and red, green and blue circles denote atoms with hybridizations of sp, sp$^2$ and sp$^3$, respectively.} 
\label{fig:casc1}
\end{figure*}

Simulation results are presented in four sections. 
In section A, representative ballistic cascades for different PKA energies are visualized and compared through displacement and cluster metrics. 
In section B we collate statistical quantities in the ballistic phase, including the kinetic energy maximum and ballistic phase length. 
In section C, statistics on the bond formation, bond breaking, cascade extent and displacements are provided. 
In section D, cross-linking in the thermalization phase is studied through tracking the number of cross-linked molecules, the kinetic energy distribution, and the temperature of the largest cluster over time. 

\subsection{\label{rsec:1}Individual Cascades}
Figure~\ref{fig:casc1} presents a representative 1~keV cascade in fullerite. 
All visualizations are produced with the blue spheres as the displacements and the solid grey lines as the trajectories, omitting all other atoms from view. 
In Fig.~\ref{fig:casc1}(a) the PKA has broken from its original molecule after 0.01~ps and is detected as a displacement. 
Between (a) and (b) the first collision has occurred, splitting the cascade into two sub-cascades.
At 0.05~ps the maximum kinetic energy is less than half its original value, producing six displacements.
Comparing (b) with (e) shows that the rightmost sub-cascade is more energetic and channels close to 50~\AA\ through the material before a collision. 
At 0.12~ps the cascade has produced a significant proportion of the displacements with only a tenth of the initial PKA energy remaining. 
Only six more displacements are produced between (f) and (g) near the end of the ballistic phase. 
Following (g), no more displacements are formed and a slow thermalization of bond breaking and formation follows. 
For further visualizations of the cascade see Supplementary Video~1 and Supplementary Video~2.

Figure~\ref{fig:casc1}(h) extends (g) by showing the underlying fullerite structure that is modified by the cascade.
Panel (h) is produced by first applying the cluster analysis tool and then omitting any cluster that contains 60 atoms. 
This automatically removes undamaged C$_{60}$ molecules, and leaves only molecules which have cross-linked (e.g. C$_{120}$), molecules which lost or gained atoms (e.g. C$_{59}$ or C$_{61}$), individual atoms or small fragments. 
By 0.28~ps, most atoms and fragments have recombined with a larger cluster.
Throughout the cascade, two main processes are seen, the cross-linking between molecules which cluster around the trajectory paths, and intermittent channeling indicated by long trajectories without cross-linking. 
Clusters of fragmented C$_{60}$ are not connected entirely along the paths in (h) due to the channeling but rather produce pockets of damaged material. 
Damage to the underlying structure results in the formation of dimers, molecular chains via sp$^3$ bridges, or amorphous material formed from direct collisions against the molecular cages. 
Notably, the cascades only rarely produce four-membered carbon rings which are commonly found in the photo-induced polymerization of fullerite~\cite{ZHOU-PHOTOINDUCED-OPT}. 

\begin{figure}[b]
    \centering
    \includegraphics[width=\linewidth]{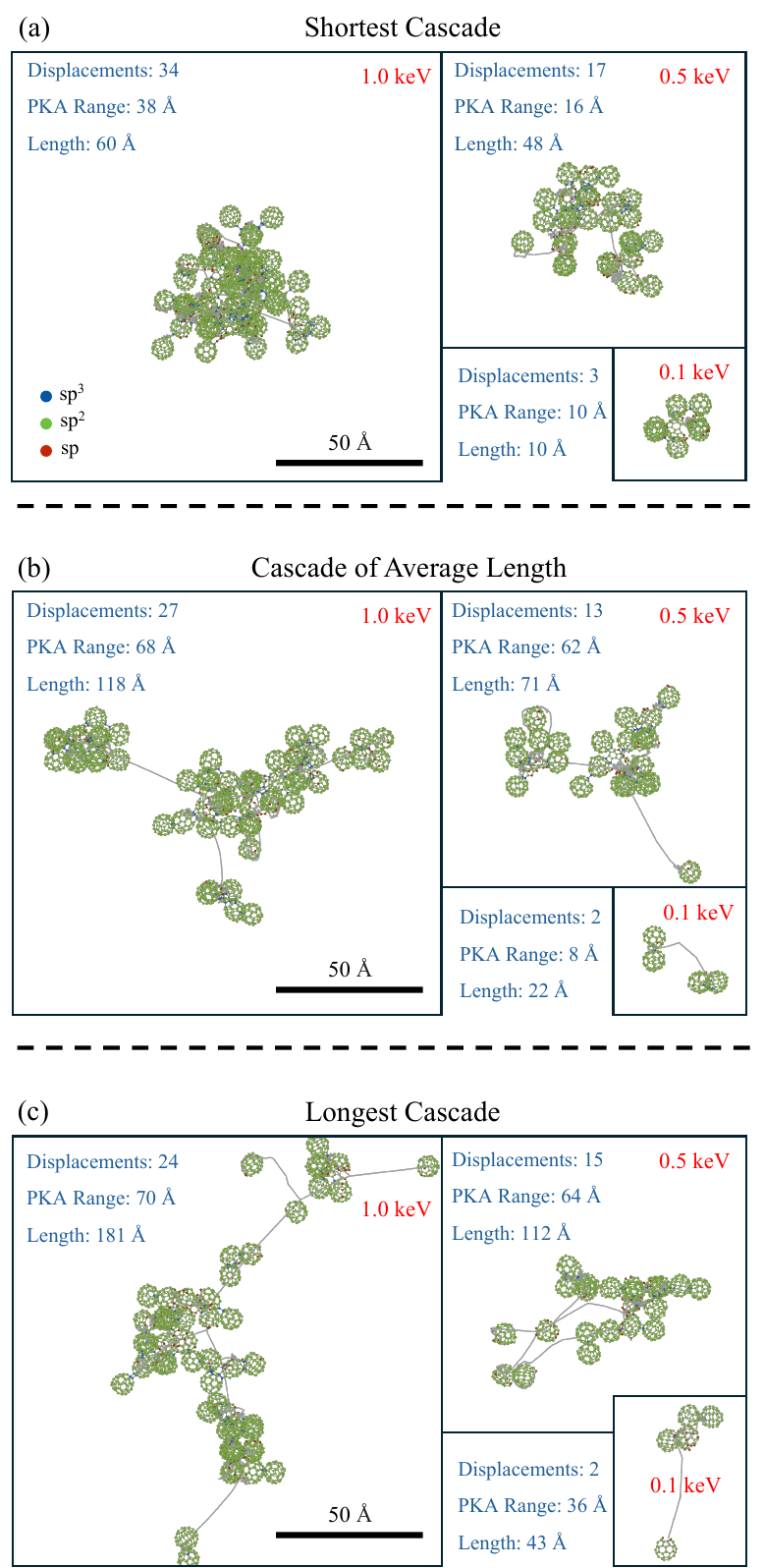}
    \caption{Representative cascades (smallest, average, greatest length) for PKA energies of 0.1, 0.5 and 1~keV. All views are along a Cartesian direction and show the cascade after 5~ps. Atoms in clusters containing 60 atoms are not displayed. Grey lines show trajectory of displacements. }
\label{fig:lengthex}
\end{figure}

While all radiation cascades have a strong stochastic signature, this is especially the case in fullerite due to its molecular nature. 
The extent of variability is highlighted in Fig.~\ref{fig:lengthex} which shows cascades at 0.1, 0.5 and 1~keV after 5~ps.
For each energy, three cascades are selected from the 25 directions: the shortest cascade, the longest cascade, and the cascade with the closest length to the average.
The short cascades with dense clustering in (a) are induced by the initial collisions of the PKA with the original C$_{60}$ cage.
Close proximity of the PKA with neighbouring atoms leads to a rapid series of high-energy collisions localized around the original molecule.
The localized clusters comprise a variety of intermolecular bonding including chains of sp carbon atoms, single or double sp$^3$ connections, and open cages which form net-like surfaces between molecules. 

Cascades with extensive channeling as seen in (c) are induced by the PKA breaking away from its original C$_{60}$ and traveling into gaps between neighbouring molecules or through the hexagonal faces of the cages.  
Pockets of damaged molecules are connected solely by the trajectory lines and extend approximately three times the length of structures in (a).
Individual C$_{60}$ cages with atom vacancies or additional atoms are commonly observed and are less frequent in the localized structures of (a). 
As the cascades in (c) are spread throughout larger volumes of space, the energy introduced by the PKA is spread across more molecules. 
This leads to a decrease in the number of displacements proportional to the increase of the cascade length from (a) to (c). 

Structures in Fig.~\ref{fig:lengthex}(b) are a midpoint between the large local clusters prevalent in (a) and the extended channeling in (c). 
Following the trajectory lines in (b), we find some displacements produce spiral paths which wrap around damaged C$_{60}$ cages.  
These paths provide examples of molecular rotation induced by collisions between the cages and the displacements. 
Note that due to the removal of the intact C$_{60}$ cages, a majority of the molecular rotation produced through indirect collisions and the ambient temperature is visually omitted.  

\begin{figure}[t]
    \centering
    \includegraphics[width=\linewidth]{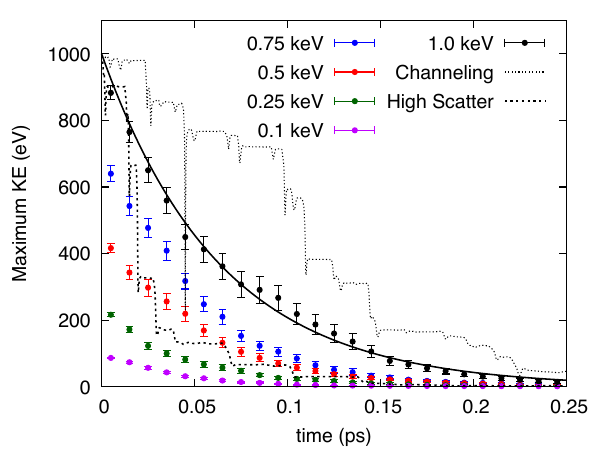}
    \caption{Time evolution of the maximum kinetic energy in cascades with differing PKA energies. Data is averaged over 25 initial directions and the error bars are the standard error in the mean. Dashed lines are 1~keV examples of a cascade with channeling and a cascade with high scatter.
    The solid line is a 1~keV exponential fit to guide the eye.
    } 
\label{fig:kemax}
\end{figure}

\subsection{\label{rsec:2}Cascade Statistics}
In our earlier work on diamond cascades \cite{BUCHAN-DIAMONDCASC} we found that the maximum kinetic energy (KE$_{max}$) was a useful metric to quantify energy transfer in the ballistic phase. Figure~\ref{fig:kemax} plots this quantity as a function of time and PKA energy.
Also shown in the figure are two edge cases at 1~keV that highlight significant variability. 
The dotted line indicates a cascade with channeling; many glancing collisions, seen by sharp downward spikes in KE$_{max}$, are required to dissipate the energy. 
The bold-dashed line shows the other extreme of behaviour, with a handful of high-energy collisions reducing KE$_{max}$ to nearly a tenth of its original value in just 0.05~ps.
Despite this variability, when averaged over the 25 PKA directions, the data for KE$_{max}$ becomes quite smooth, and the effect of the PKA energy is easily seen.

In our previous work~\cite{BUCHAN-DIAMONDCASC} we fit the KE$_{max}$ data to a decaying exponential of the form
\begin{equation*}
\textrm{KE}_{max} = E_{0}\exp(-t/t_{0})    
\end{equation*}
where $E_{0}$ is the initial PKA energy and $t_{0}$ is a time constant.
To create a comparison to fullerite, the 1~keV cascades in Fig.~\ref{fig:kemax} are fit in the same manner and visualized as the solid black line.
The time constant of fullerite is 0.064~ps, four-times that of diamond at 0.016~ps.
These contrasting timescales reflect the difference in structural topology, with the dense network of sp$^3$ bonds in diamond yielding a quick thermal dissipation of the atoms compared to the weakly-interacting molecules in fullerite. 

\begin{figure}[t]
    \centering
    \includegraphics[width=\linewidth]{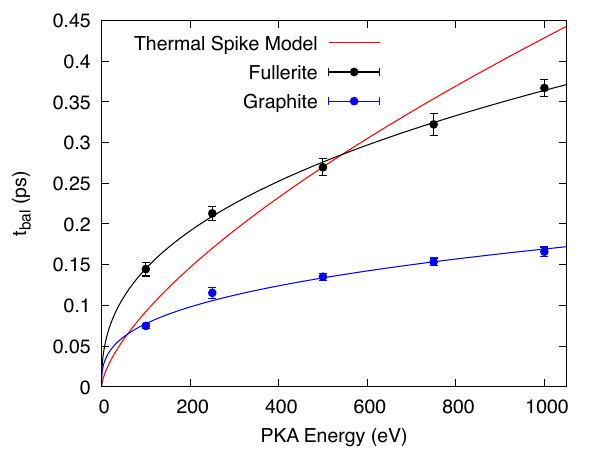}
    \caption{Duration of the ballistic phase as a function of PKA energy. The duration is defined as the time when the kinetic energy maximum falls below half the strength of a sigma bond (1.84~eV).  
    Error bars are the standard error in the mean and solid lines depict the power-law fits with the form $a$E$^{b}$. 
    Exponents are 0.34, 0.40 and $\frac{2}{3}$ for graphite, fullerite and the spherical thermal spike model respectively. 
    Graphite data is derived from our previous study~\cite{CHRISTIE-GRAPHITECASC}.}
\label{fig:tbal}
\end{figure}

Having analysed the dispersion of kinetic energy in fullerite cascades, we shift our focus to the timeframe of the ballistic phase.  
The duration of the ballistic phase is quantified by the instant at which the KE$_{max}$ first drops below half the strength of a sigma bond (1.84~eV) and corresponds roughly to the time where no more displacements are produced.   
Fig.~\ref{fig:tbal} depicts this duration as a function of the PKA energy for fullerite and graphite.
The solid lines are power-law fits ($aE^{b}$) which are first introduced in our study of graphite~\cite{CHRISTIE-GRAPHITECASC} and provide a good approximation of the weak energy dependence in the ballistic phase duration for both materials. 
The longer ballistic phase of fullerite is clear in the divergence of the two curves at increasing PKA energy, with graphite having an exponent of 0.34 and fullerite having an exponent of 0.40. 
Cascades in metals and oxides often produce a local melting around collision sites followed by a fast cooling and recrystallization phase, namely a thermal spike. 
In these systems a simple model of heat diffusion \cite{MARKS-1997-THERMALSPIKE} predicts a power-law character of the energy with an exponent of $\frac{2}{3}$ (red line in Fig.~\ref{fig:tbal}). 
With an exponent of 0.40, fullerite is far from a thermal spike, albeit less so than graphite whose strong bonds and extended sheets rapidly disperse energy.

\begin{figure}[t]
    \centering
    \includegraphics[width=0.98\linewidth]{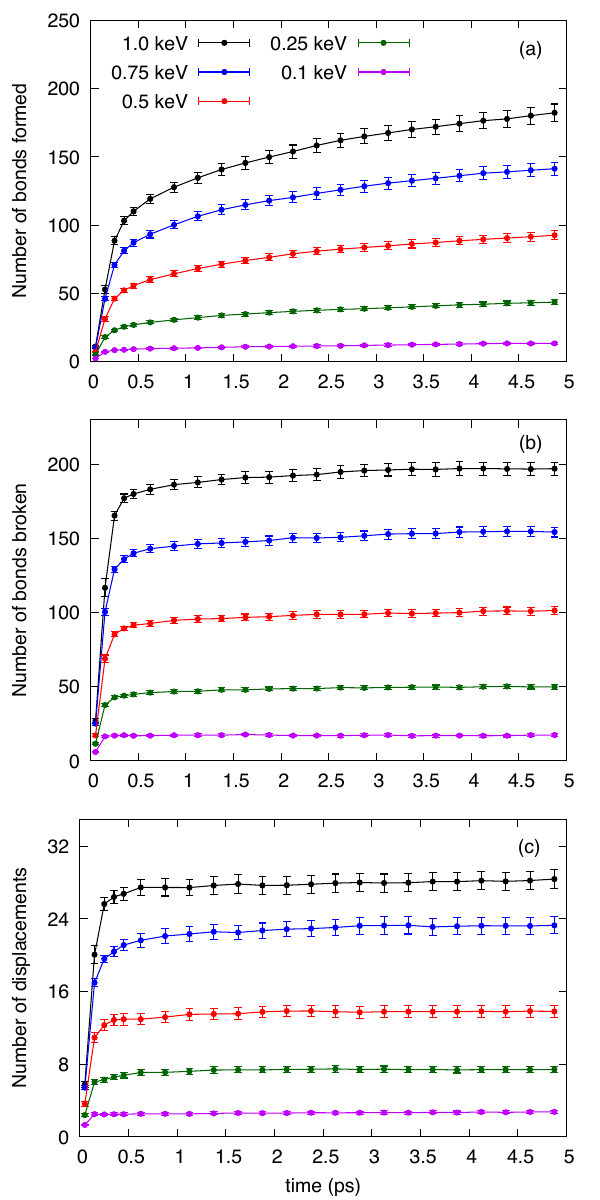}
    \caption{Bonds formed, bonds broken and number of displacements as a function of time for the first 5~ps. Data is averaged over the 25 PKA directions and error bars denote the standard error in the mean. The abrupt change in slope in all three graphs highlights the transition between the ballistic and thermalization phases.}
\label{fig:bonds}
\end{figure}

\subsection{\label{rsec:3}Bonding \& Displacements}
\begin{figure}[t]
    \centering
    \includegraphics[width=\linewidth]{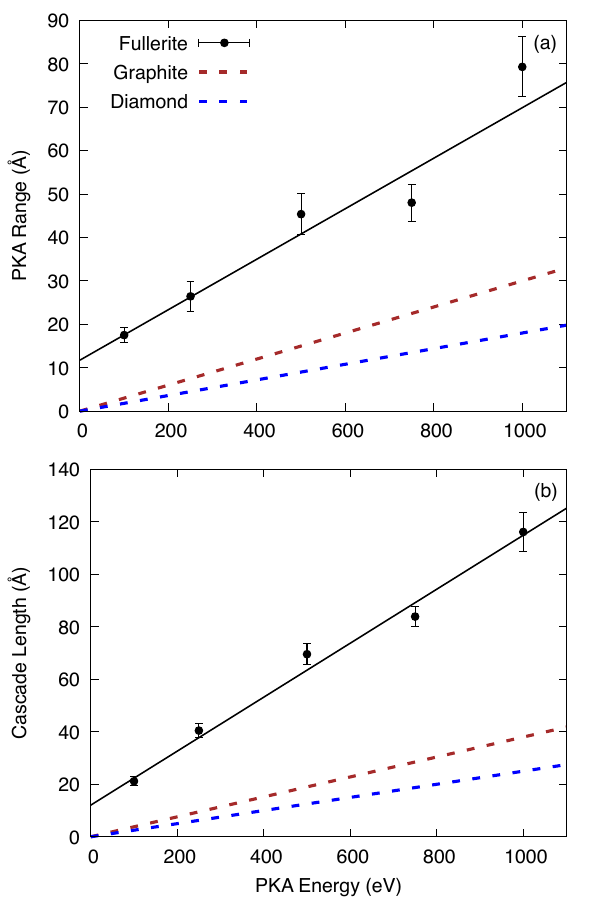}
    \caption{PKA range (a) and cascade length (b) as a function of PKA energy. Data points are averaged over 25 PKA directions and the error bars are the standard error in the mean. Solid lines are linear fits with gradients of 58~\AA\//keV and 103~\AA\//keV. The graphite and diamond data is from our previous work~\cite{BUCHAN-DIAMONDCASC}.}
\label{fig:casclength}
\end{figure}

An analysis of radiation damage in fullerite is presented through three metrics in Fig.~\ref{fig:bonds}, with (a) depicting the number of bonds formed, (b) depicting the number of bonds broken and (c) showing the number of displacements produced as a function of time. 
Two distinct slopes appear in all metrics, a sharp increase in displacements and bond changes within the first 0.5~ps, followed by either a slow increase in bond formation and destruction, or a plateau of the displacement production. 
Previously we presented the ballistic phase duration in Fig.~\ref{fig:tbal} which ranged from 0.15 to 0.35~ps for PKA energies of 0.1 to 1~keV. 
This matches well with the sharp initial slopes, representing damage to the structure via rapid collisions. 
Following the ballistic phase, thermal motion dominates and a slow process of bond formation is seen through the steady slope after 0.5~ps. 
The standard deviation in the slope of (a) increases with time and reaches a value of 32 bonds formed at 4.9~ps, reflecting the stochastic nature of the cascades. 

Figure~\ref{fig:casclength} shows the average range of the PKA and the length of the cascades as a function of PKA energy. Data is computed after 5~ps has elapsed by which point the ballistic phase is well and truly complete.
Both quantities are well-described via a linear fit (solid black lines) with a non-zero y-intercept.
The y-intercepts in panel (a) and (b) are 11.7 and 12.1~\AA, respectively, which are roughly similar to the diameter of the C$_{60}$ molecules (7.378~\AA) and might be related to the distance an atom moves during free rotation. 
Gradients of the PKA range and cascade length are 58~\AA\//keV and 103~\AA\//keV, respectively, approximately triple the PKA range and cascade length of graphite and diamond.
As a consequence, the volume of a simulation box that contains a fullerite cascade is at least 27 times that of graphite or diamond at the same PKA energy.
Extrapolating the cascade length to a 2.5~keV PKA in fullerite (the maximum energy in our previous studies) would require a box length close to 300~\AA\ or 2 million atoms.
We decided to limit the PKA energy to 1~keV in this study due to the high computational cost associated with a significant number of atoms. 

\begin{figure}[t]
    \centering
    \includegraphics[width=\linewidth]{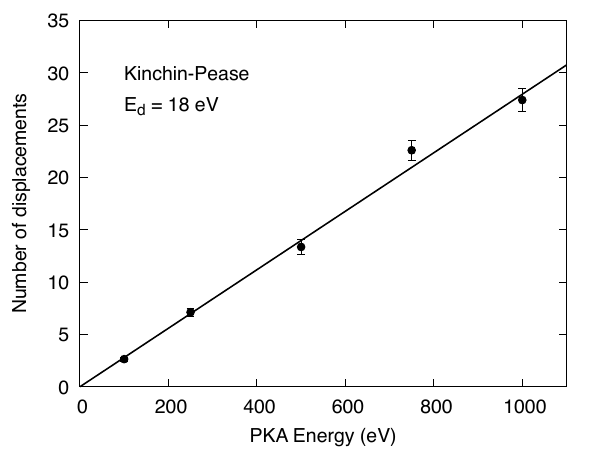}
    \caption{Number of displacements as a function of PKA energy. Data points are averaged over 25 PKA directions and error bars are the standard error in the mean. The fit line has a slope of 0.028~atoms/eV and corresponds to a Kinchin-Pease E$_d$ of 18~eV. }
\label{fig:disp}
\end{figure}

\begin{figure*}[t]
    \centering
    \includegraphics[width=\linewidth]{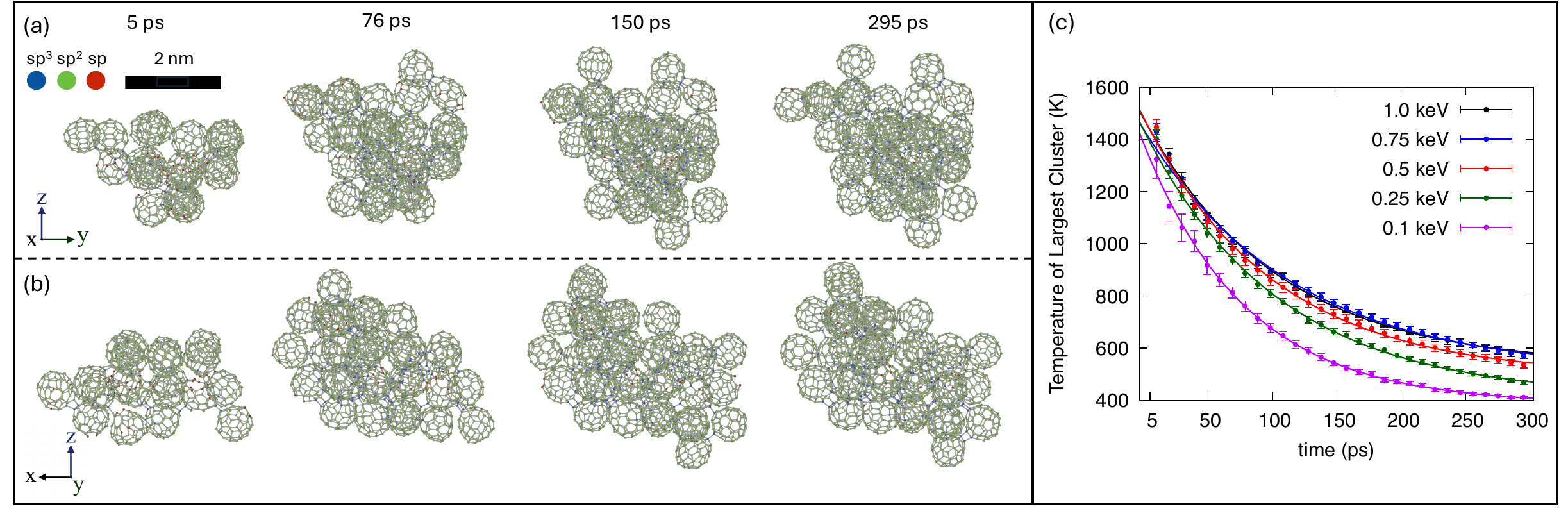}
    \caption{ (a-b) Visualization of growth in the largest cluster of a 1~keV cascade over 290~ps. (a) viewing along the x-axis and (b) viewing along the y-axis. (c) Mean temperature of the largest cluster as a function of time for 0.1, 0.25, 0.5, 0.75 and 1.0~keV PKA cascades. Data is averaged over 25 PKA directions and error bars are the standard error in the mean. Exponential fits are plotted to guide the eye. }
\label{fig:clustemp}
\end{figure*}

The radiation resistance of fullerite can be characterized by plotting the number of displacements as a function of the PKA energy, as shown in Fig.~\ref{fig:disp}. 
A simple model to determine the threshold displacement energy of a material was derived by Kinchin and Pease through approximating the cascades as collisions of hard spheres~\cite{KP}.
The black line is a linear fit through the origin and can be related to the Kinchin-Pease (KP) equation,
\begin{equation*}
N_{d} = \frac{E_{PKA}}{2E_d}
\end{equation*}
where $N_{d}$ is the number of displacements, $E_{PKA}$ is the energy of the PKA and $E_d$ is the threshold displacement energy. 
The fit yields a gradient of 0.028~atoms/eV which corresponds to a threshold displacement energy of 18~eV. 
For comparison, experimental and theoretical values of $E_{d}$ in fullerite range from 10 to 24.1~eV~\cite{FULLER-1996-Td, PARILIS-1994-Td,STOCKETT-2018-Td} and molecular dynamics studies of C$_{60}$ provide values ranging from 15 to 29~eV~\cite{STOCKETT-2018-Td, TOMITA-2002-MD-Td, CUI-1994-MD-Td}; our findings are consistent with both ranges.
We note that graphite has a similar E$_d$ of 25~eV when calculated via the same KP model~\cite{CHRISTIE-GRAPHITECASC}; this is reasonable since both materials have comparable sp$^2$ bonding. 

\subsection{Thermalization Phase Statistics \label{sec:thermal}}
In Fig.~\ref{fig:bonds} we showed that the bond formation proceeded beyond 5~ps in the thermalization phase, albeit at a much slower rate than the ballistic phase. 
Fig.~\ref{fig:clustemp}(a-b) shows the largest cluster in a 1~keV cascade over 290~ps, where all other molecules are omitted. 
Looking down the x-axis, Fig.~\ref{fig:clustemp}(a) depicts the initial size of the cluster to be approximately 2~nm along the z-axis and grows a further 3~nm over 290~ps. 
A majority of the growth occurs in the first 100~ps and is driven by thermal motion within the cluster.
The primary process for dissipating the kinetic energy introduced by the PKA is through the creation of new sp$^3$ cross-links both within the cluster and by connecting two neighbouring molecules. 
To study the dissipation of kinetic energy from a cluster, the temperature of the largest cluster in the material is tracked over time in Fig.~\ref{fig:clustemp}(c). 
Exponential fits are introduced with a non-zero y-offset to guide the eye.  
Beginning a few picoseconds into the thermalization phase, the cluster temperature is approximately 1500~K independent of the PKA energy. 
These clusters are clearly not at equilibrium and interact via an annealing process rather than a thermal spike, where the energy is transferred through slow-forming molecular defects rather than interconnected metal or oxide lattices. 
Over 150~ps all cascades have dropped below half the maximum temperature with 0.1~keV clusters depicting the most significant decrease, followed by 0.25, 0.5, 0.75 and 1.0~keV clusters.

\begin{figure}[t]
    \centering
    \includegraphics[width=\linewidth]{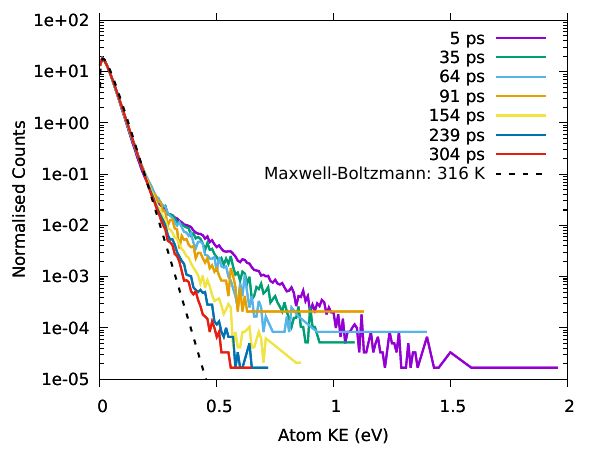}
    \caption{Distribution of the kinetic energy of atoms as a function of time for a 1~keV cascade. The dashed black line is the Maxwell-Boltzmann distribution for the system at equilibrium. }
\label{fig:kedistr}
\end{figure}

To get a total picture of the slow thermalization process, the kinetic energy of each atom across all 1~keV cascades is histogrammed into a distribution in Fig.~\ref{fig:kedistr}.
Several time snapshots are provided from 5 to 304~ps, distinguished by colour. 
The extended thermalization phase is represented by the tail of the distributions which reduces from a maximum of 2~eV atoms to below 0.75~eV by the end of the simulations.
For reference, a Maxwell-Boltzmann distribution is plotted as a dotted-line, representing the equilibrium distribution of the kinetic energy when introducing a 1~keV PKA at 300~K. 
The convergence of the high-energy tail towards this equilibrium is on the order of hundreds of picoseconds, which highlights the significance of the low thermal conductivity of the molecular solid in the thermalization process. 
Similarly, graphite has weak thermal coupling between sheets and comparable sp$^2$ bonding, however, the heat dissipation is orders of magnitude faster due to a significantly higher number of bonds per sheet than a C$_{60}$ molecule. 

\begin{figure}[t]
    \centering
    \includegraphics[width=0.98\linewidth]{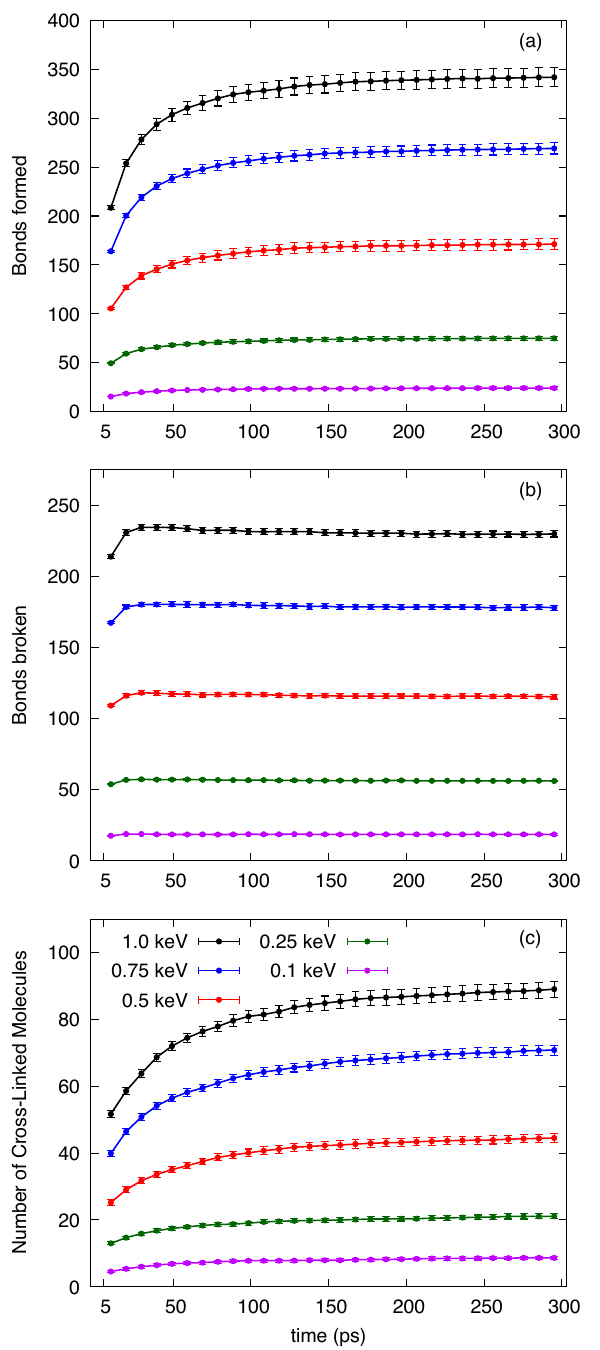}
    \caption{Number of bonds formed, number of bonds broken and number of cross-linked molecules as a function of time between 5 and 300~ps of the cascade. Data is averaged over the 25 PKA directions and error bars denote the standard error in the mean.}
\label{fig:thermalbonds}
\end{figure}

The breaking and formation of bonds in the thermalization phase is plotted over time in Fig.~\ref{fig:thermalbonds}(a) and (b), respectively. 
The two processes act on different timeframes with the breaking of bonds finishing by approximately 50~ps and the bond formation lasting for approximately 200~ps. 
As the PKA energy decreases from 1.0 to 0.1~keV the scale of the error bars decreases until the value is contained within the points displayed. 
The stochastic nature of the cascades and the thermalization process is evident with some PKA directions differing in the number of bonds by up to 30\% of the total value. 
To track the topology of the cascades statistically, Fig.~\ref{fig:thermalbonds}(c) shows the number of cross-linked molecules formed during the thermalization phase. 
The metric is calculated by omitting all clusters from the data that contain small fragments (e.g. C$_{1}$, C$_{2}$, C$_{3}$) or are similar in size to the undamaged molecules (e.g. C$_{59}$, C$_{60}$, C$_{61}$).
This is completed through introducing a lower cutoff of the cluster size, chosen to be 100, and the remaining clusters are divided by 60 to approximate the number of cross-linked molecules.    
In both the 0.1 and 0.25~keV cascades the error in the mean is comparable to the visual point size, and a majority of the molecules for all PKA energies are cross-linked in the first 100~ps. 
This trend in the formation of cross-linked molecules matches the growth of the 1~keV cluster seen in Fig.~\ref{fig:clustemp}(a-b). 

\begin{figure}[t]
    \centering  \includegraphics[width=\linewidth]{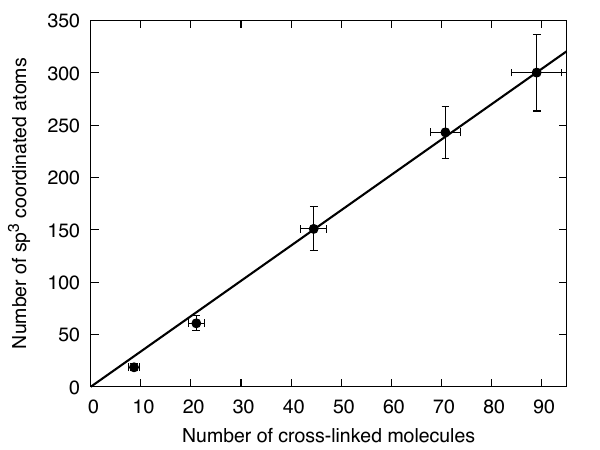}
    \caption{Number of sp$^{3}$ atoms as a function of the number of cross-linked molecules at the end of the simulations. Error bars signify the 95\% confidence interval along both axes. The solid line is a linear fit with a gradient of 3.4~sp$^3$/molecule.}
\label{fig:clustersp3}
\end{figure}

Motivated by the high computational expense of the cluster calculation, alternative metrics are required for simulations with a significant number of atoms (i.e. fullerite cascades with PKA energies larger than 1~keV).
A natural proxy for cross-linked molecules are sp$^3$ coordinated atoms, which are formed by the connection of sp$^2$ atoms on two different C$_{60}$ cages via thermal motion or atomic collisions.
In terms of computational cost, the coordination of each atom is simple to compute with a neighbour list compared to a depth-first search for clusters. 
As shown in Fig.~\ref{fig:clustersp3}, the number of sp$^3$ atoms has a linear dependence on the number of cross-linked molecules. 
One data point is plotted for each PKA energy and the error bars denote the 95\% confidence interval for both metrics. 
Using the linear fit (solid line), a gradient of 3.4~sp$^3$/molecule suggests that cross-linked molecules are connected using three or four cross-links. 
Notably, the characteristic sp$^3$ cross-links produced by the cascades are amorphous links in comparison to the ordered connections formed from the pressure-induced polymerization~\cite{SUNDQVIST-2004} or the photo-polymerization~\cite{ZHOU-PHOTOINDUCED-OPT} of fullerite.  

\section{Discussion \& Conclusion}
One of the main insights of this study is the extensive thermalization phase of cascades in a molecular solid compared to most crystalline solids such as graphite, diamond, metals or oxides. 
All cascades in fullerite exhibit the cross-linking of the C$_{60}$ cages through direct collisions with the PKA, which deposits the energy into small molecular clusters.
Hobday~\textit{et al}.~\cite{HOBDAY1997-AR-FULLERITE} conducted a similar MD study of fullerite thin-films in 1997 and observed cross-linking and polymerisation of the C$_{60}$ cages, but didn't consider the extensive thermalization phase as the surface sputtering was the primary concern, and the simulations were followed for only 5~ps. 
Later work by Czerwinski~\textit{et al}.~\cite{CZERWINSKI-ARN-C60-FULLERITE} used large clusters of atoms (Ar$_{n}$ clusters and C$_{60}$ molecules) as the projectiles but also didn't mention the thermalization phase, likely due to the computational limit of simulating large fullerite systems of 1.3--2.4 million atoms.
The surface impact study by Shinya \textit{et al}.~\cite{SHINYA2009-C02-FULLERITE} suggested a long thermalization timescale of the fullerite surface with a duration of at least 100~ps from the bombardment of (CO$_2$)$_n$ clusters at energies of 5--14 keV.   
We observe in our simulations that a cross-linking process both within and between the clusters is the predominant mechanism for thermal dissipation during thermalization, while contributions from the weak van der Waals forces are negligible in comparison. 
Shinya \textit{et al}. also suggested this mechanism and found the molecular clusters in the cascades retain temperatures of over 1000~K in the first 100~ps.
We also observe this effect, (see Fig.~\ref{fig:clustemp}) and further emphasize that the clusters maintain a temperature above equilibrium for hundreds more picoseconds.
In contrast, most crystalline solids exhibit thermalization phases that last on the order of a few picoseconds due to their comparatively larger thermal conductivity~\cite{CHRISTIE-GRAPHITECASC,BUCHAN-DIAMONDCASC,TIO2}.
Theoretically, since the duration of the thermalization phase in fullerite is dependent on the bond formation, an increase in the flux of PKA projectiles would increase the number of cross-links and, in turn, increase the thermal conductivity of the solid.
Further studies of radiation damage in fullerite could test the relationship between the duration of the thermalization phase and the flux of the PKA projectiles. 

The cascade dynamics in fullerite are unique, as channeling is more prominent than in graphite or diamond and leads to an extensive cascade length, subtle deviations in the KE$_{max}$, and a longer ballistic phase duration. 
While the difference in cascade dynamics between fullerite and diamond is understandable, with loosely connected molecules compared to numerous interconnected sp$^3$ bonds that resist the radiation damage, it is not so obvious why graphite doesn't produce similar dynamics.
In particular, cascades in fullerite travel 103~\AA/keV compared to 38~\AA/keV in graphite or 25~\AA/keV in diamond; despite the similar sp$^2$ bonding and van der Waals forces, graphite is closer to diamond than fullerite. 
There have been few studies on the channeling in fullerite, one theoretical study for ion, positron and electron channeling by Zhevago and Glebov~\cite{Channeling1} looked at the $\langle$100$\rangle$ and $\langle$110$\rangle$ directions, which produced stable channeling for ions and have potential wells of 7~eV in the xy-plane and 14~eV between the planes. 
Other channeling events include atoms traveling through the hexagonal rings of the C$_{60}$ cages; the straight trajectory line in Fig.~\ref{fig:casc1}(c) is an example of this process.  
Notably, a study into the defects of graphite by Kaxiras and Pandey~\cite{Graphite-Hexagonal-Defect} calculated the energy of a defect in a hexagonal ring to be 19.5~eV, this suggests a similar barrier strength for an atom to pass through the hexagonal faces of C$_{60}$. 
Graphite also has extensive channeling with common paths parallel to the c-axis or pseudo-channels that pass through hexagonal faces ($\langle0001\rangle$ and $\langle10\bar{1}2\rangle$ as seen in our previous study~\cite{CHRISTIE-GRAPHITECASC}). 
The longer cascade length of fullerite may be attributed to the lower energy barriers for channeling between the molecules ($\sim$7--14~eV) compared to the energy barrier required to channel through a graphene layer ($\sim$19.5~eV).

A benefit of the metrics used in this paper is their application in future studies of fullerite. 
Extrapolating the fit of the cascade length to higher energy PKAs can provide an estimate of the size of the simulation cell which contains the cascades. 
The kinetic energy distributions (in 1~keV cascades) approach equilibrium after 300~ps, which provides a lower bound for the thermalization phase duration. 
For efficient use of computer time, a rough cutoff of 100~ps is reasonable in 1~keV simulations as most bond/cross-link formation occurs in this period. 
The threshold displacement energy, number of bonds made/broken and number of cross-linked molecules can provide insight into the build-up of radiation damage in fullerite and guide ion implantation experiments using 0.1--1~keV projectiles.  

In summary, we have performed molecular dynamics simulations of a carbon PKA in bulk fullerite. 
Cascades in fullerite combine channeling with high scattering processes to produce a wide variety of structures from disconnected small pockets of damage to dense regions of cross-linked molecules.  
It is made clear that fullerite, as a molecular solid, has a thermalization phase orders of magnitude longer than most crystalline solids.
The threshold displacement energy is estimated using the Kinchin-Pease method and is consistent with literature values. 
Important quantities such as the kinetics, bond connectivity, cross-linking and cascade length characterize the radiation damage in fullerite and provide an entry point for further simulations. 

\section*{Acknowledgments}
EPT acknowledges support through an Australian Government Research Training Program Scholarship. 
This work was supported by resources provided by the Pawsey Supercomputing Research Centre’s Setonix Supercomputer, with funding from the Australian Government and the Government of Western Australia.
This research was funded by the Australian Research Council (ARC) through project DP230100231.

\bibliography{ref}
\end{document}